\documentclass[11pt,a4paper]{article}
\pdfoutput=1
\usepackage{jheppub}
\usepackage[utf8x]{inputenc}
\usepackage{graphicx}
\usepackage{subfig}

\begin{document}

\title{Axial Hall effect and universality of holographic Weyl semi-metals}
\author[1]{Christian Copetti%
\note{email: christian.copetti@uam.es}}
\author[2]{, Jorge Fern\'andez-Pend\'as%
\note{email: j.fernandez.pendas@csic.es}}
\author[3]{, Karl Landsteiner%
\note{email: karl.landsteiner@uam.es}}
\affiliation{Instituto de F\'isica Te\'orica UAM/CSIC,\\c/ Nicol\'as Cabrera 13-15, Cantoblanco, 28049 Madrid, Spain}
\date{Dated: \today}
\abstract{
The holographic Weyl semimetal is a model of a strongly coupled topological semi-metal.
A topological quantum phase transition separates a topological phase with non-vanishing 
anomalous Hall conductivity from a trivial state. We investigate how this phase transition
depends on the parameters of the scalar potential (mass and quartic self coupling) finding
that the quantum phase transition persists for a large region in parameter space. We then compute the axial Hall conductivity. The algebraic structure of the axial anomaly predicts
it to be $1/3$ of the electric Hall conductivity. We find that this holds once a non-trivial
renormalization effect on the external axial gauge fields is taken into account.
Finally we show that the phase transition also occurs in a top-down model based on a 
consistent truncation of type IIB supergravity.
}
\keywords{Holography, Weyl Semimetals, Quantum Phase Transitions}
%\arxivnumber{1234.5678v2}
\preprint{IFT-UAM/CSIC-16-132}
\maketitle
\flushbottom
\section{Introduction}\label{sec:intro}
A Weyl semimetal is a topological state of matter with chiral fermions as electronic quasiparticles \cite{WSM1,WSMreviewTV}. The exotic electronic transport properties of such systems can be understood as effects due to chiral anomalies \cite{Landsteiner:2016led}. 
For instance the chiral magnetic effect
\cite{Fukushima:2008xe} gives rise to an enhanced
conductivity along an external magnetic field \citep{Li:2014bha} with characteristic
quadratic scaling in the magnetic field strength.

Our motivation for investigating the physics of Weyl semimetals via holographic models is twofold. First the Fermi-velocity in Weyl semimetals is low compared to the speed of light and this might give rise to an effective large fine structure constant similar as in Graphene. In this case inherently strongly coupled models based on holography might give a good description of the physics similar to what happens in the strongly coupled quark 
gluon plasma. Secondly holography has been extremely useful in discovering and understanding
anomalous transport phenomena. So it is natural to explore holographic models of Weyl semimetals and their anomaly related transport properties 
\cite{Landsteiner:2015pdh,Landsteiner:2016stv,Grignani:2016wyz,Ammon:2016mwa}. Anomalies are universal
properties of quantum field theories and independent of the coupling constant. Weakly coupled and strongly coupled models have the same basic anomaly
determined transport coefficients. In weakly coupled field theory one needs to
deal with finite but undetermined one-loop diagrams\cite{Jackiw:1999qq}. In holography these subtleties of the regularization procedure are mapped onto the classical gravitational dynamics in the fifth dimension. Furthermore, while it is true that the triangle diagram of the chiral anomaly
does not get contributions beyond one loop this is not true for the external fields that
have to be attached to the vertices or equivalently for the charges that are the
couplings of the external fields to the currents sitting at the vertices. 
All these issues arise also in holography. But it is much easier to 
understand the physics and fix the ambiguities due to the fact that the anomaly appears in
the classical equations of motion via Chern-Simons terms. In holography the ambiguities appear as boundary
Chern-Simons (counter-)terms. Demanding gauge invariance fixes them  uniquely.
Field renormalization appears as the dynamics in the fifth dimension, the holographic RG flow.
Precisely such a field or charge renormalization plays a major role in the present paper.

We start out in section two with an investigation of the parameter space of the holographic
Weyl semimetals introduced in \cite{Landsteiner:2015lsa,Landsteiner:2015pdh}\footnote{For alternative
holographic approaches to the physics of Weyl semimetals see \cite{Gursoy:2012ie, Jacobs:2015fiv, Hashimoto:2016ize}}. 
This is  a comparatively simple bottom-up model 
in which the symmetries and breaking patterns are the guidelines to write down an effective
Lagrangian for a holographic model. It depends on a variety of parameters in the 
scalar potential such as the mass and the quartic self coupling. We study how the quantum
phase transition between the topological phase and the trivial phase depends on the bulk mass, the quartic self coupling and the charge of the scalar field. Generally the phase transition is still present but for fixed bulk mass there is an upper
bound on the quartic self coupling above which the trivial phase can not be reached
anymore. Similarly for fixed quartic coupling but varying the bulk mass we
find that near $m^2 L^2 = 0$, i.e. when the operator breaking the axial symmetry becomes marginal, the trivial phase becomes unreachable.  The precise value
depends on the charge of the scalar field but is always close to marginality. 

In section three we compute the axial Hall conductivity. The usual Hall conductivity is
the response in the electric current transverse to an applied electric field and in a Weyl semimetal
it has an anomalous contribution proportional to the anomaly triangle with one axial current
and two vector currents, the mixed $U(1)_A U(1)_V^2$ anomaly. 
There is however also a cubic $U(1)_A^3$ purely axial anomaly and
the algebraic structure of the anomaly fixes the strength to be precisely
$1/3$ of the AVV anomaly. At weak coupling this can be thought of as arising
from the additional symmetry factor on the triangle diagram. 
For the axial Hall conductivity, i.e. the axial current induced by a transverse
axial electric field, this predicts the value to be precisely $1/3$ of the 
electric Hall conductivity.  
The non-renormalization theorem of the anomaly guarantees this to hold
independent of the strength of the coupling constant. However if one naively computes and
interprets the result of the axial Hall conductivity this seems to hold only for the
trivial case in which the $U(1)_A$ symmetry is not broken by tree level terms
and to be at odds with the non-renormalization theorem. The resolution is that while
the triangle diagram by itself does not suffer renormalization  the external fields (or couplings) with which we can probe the anomaly can and in fact do get renormalized. In the
non-trivial topological phase of our model the axial gauge fields couple to the bulk scalar field and get renormalized along the holographic RG flow. In the infrared region of the model
only a fraction of the UV source that we apply at the boundary arrives. Much of the
axial gauge field gets screened in the scalar field background. Once we account for this
RG flow by renormalizing the coupling of the axial gauge field to the anomaly triangle
we find that the relation of axial to electric Hall conductivity is precisely $1/3$.
This factor is a property of the theory and should hold for arbitrary states.
We check this by showing that it holds indeed also at finite temperature. In the trivial
phase the axial gauge field is completely screened, there are no degrees of freedom in
the IR that could couple to them and thus the anomaly cannot be probed anymore\footnote{In
a tight binding lattice model an ever stronger RG effect was observed: what appear as
lattice deformations in the UV can couple to the chiral quasiparticles in the IR as axial gauge fields \cite{Cortijo:2016yph}}.

Finally in section four we discuss a string theory derived model. It is based on
a consistent truncation of type IIB supergravity and has been used before in the study of
holographic superconductors \cite{Arean:2010wu}. In this model there is no freedom to choose parameters in the
scalar potential or elsewhere.
We show that the topological quantum phase transition takes place also in this model 
and that the value of the axial Hall conductivity shows the same renormalization effect.

\section{Universality of the topological quantum phase transition}\label{sec:qpt}

The holographic model of Weyl semi-metals \cite{Landsteiner:2015pdh} has been shown to allow for a topological quantum phase transition at zero temperature. It interpolates between a topological phase with a nonzero Hall conductivity and a trivial phase in which the Hall conductivity vanishes. At finite temperature this
phase transition becomes a smooth crossover. 

Our aim for this section is to determine the universality and the possible extensions of this behavior. Particularly, we will try to understand how the change in the parameters of the model will affect the  phase transition and how the value
of the critical point varies.

The action of the model is:
\begin{equation}\label{action}
\begin{aligned}
S %_g 
= &\int d^5 x \sqrt{-g}\left\{\frac{1}{2\kappa^2} \left(R %-2\Lambda_{AdS_5}
+ \frac{12}{L^2}\right) -\frac{1}{4}F^2 -\frac{1}{4}F_5^2 -(D_M \Phi)^* D^M \Phi -V(\Phi) + \right. \\
&+\left. \frac{\alpha}{3} \epsilon^{MNOPQ}A_M\left(F^5_{NO}F^5_{PQ} +3 F_{NO}F_{PQ}\right)\right\},
\end{aligned}
\end{equation}
where $\kappa$ is the gravitational coupling constant and $L$ is the AdS length. From now on we set $2\kappa^2=L=1$. Variation of the action gives the following equations of motion:
\begin{align}
&\begin{aligned}
0 =& G_{MN} - \Lambda g_{MN} + \frac{1}{8} g_{MN} F_{RS} F^{RS} - \frac{1}{2} {F_M}^Q F_{NQ} + \frac{1}{8} g_{MN} F_{RS}^5 F_5^{RS} - \frac{1}{2} F_M^{5\ Q} F_{NQ}^5 \\
& + \frac{1}{2} g_{MN} \left[ (D_P \Phi )^* D^P \Phi + m^2 \Phi^* \Phi + \frac{\lambda}{2} ( \Phi^* \Phi )^2 \right] - \frac{1}{2} ( D_M \Phi )^* D_N \Phi - \frac{1}{2} ( D_N \Phi )^* D_M \Phi ,
\end{aligned}\\
&0 = \nabla_M F^{MS} + 2 \alpha \epsilon^{SNPQR} F_{NP}^5 F_{QR} , \label{eomvector}\\
&0 = \nabla_M F_5^{MS} + \alpha \epsilon^{SNPQR} \left( F_{NP}^5 F_{QR}^5 + F_{NP} F_{QR} \right) + i q \left( - \Phi^* D^S \Phi + ( D^S \Phi )^* \Phi \right) , \label{eomaxial}\\
&0 = \nabla_M ( D^M \Phi ) - i q A^M D_M \Phi - m^2 \Phi - \lambda (\Phi^* \Phi) \Phi ,
\end{align}
where $\nabla_M$ is the gravitational covariant derivative. We now comment on the field content of the model.

We introduce a vector gauge field $V_\mu$ and an axial gauge field $A_\mu$, whose field strengths are respectively $F = dV$ and $F_5 = dA$, in order to account for the electromagnetic and axial $U(1)$ symmetries of the system. The scalar field $\Phi$ is only charged with respect to the axial field via the covariant derivative $D_M = \partial_M - i q A_M$ and has a
quartic potential $V(\Phi)= m^2 |\Phi|^2 + \frac{\lambda}{2}|\Phi|^4$. Explicit breaking of the axial $U(1)$ symmetry
will be achieved by switching on the non-normalizable mode. 

The final ingredient in the action required to make the axial symmetry anomalous is the five-dimensional Chern-Simons term with the particular choice for the coefficients shown above.
They are chosen such that the gauge variation of the action 
mimics the VVA and AAA anomalies of Dirac fermions and preserves the vector
like gauge symmetry.
It allows the definition of a conserved consistent\footnote{We denote consistent currents throughout the paper with calligraphic letters $\mathcal{J}$ to distinguish them from covariant ones.} vector current and an anomalous consistent axial current:
\begin{align}
 \mathcal{J}^\mu &= \lim_{r \to \infty} \frac{\delta S}{\delta V_\mu} = \lim_{r \rightarrow \infty} \sqrt{-g} \left( F^{\mu r} + 4 \alpha \epsilon^{r \mu\nu\rho\sigma} A_\nu F_{\rho\sigma} \right), \label{vectorcurrent} \\
 \mathcal{J}_5^\mu &= \lim_{r \to \infty} \frac{\delta S}{\delta A_\mu} = \lim_{r \rightarrow \infty} \sqrt{-g} \left( F_5^{\mu r} + \frac{4 \alpha}{3} \epsilon^{r \mu\nu\rho\sigma} A_\nu F_{\rho\sigma}^5 \right), \label{axialcurrent}
\end{align}
which, by the use of the radial component of (\ref{eomvector}) and (\ref{eomaxial}), satisfy the following conservation relations:
\begin{align}
\partial_\mu \mathcal{J}^{\mu} &= 0 \\
 \partial_\mu \mathcal{J}_5^\mu &= - \frac{\alpha}{3} \epsilon^{\mu\nu\rho\sigma} \left( F_{\mu\nu} F_{\rho\sigma} + 3 F_{\mu\nu}^5 F_{\rho\sigma}^5 \right) + i q \sqrt{-g} \left( \Phi^* D^r \Phi - (D^r \Phi)^* \Phi \right).
\end{align}
By the identification of the consistent anomaly for the axial current with that of a free massless Dirac fermion we can identify $\alpha = \frac{N^A_f N_c^2}{16 \pi^2}$ where $N^A_f$
is the number of Dirac flavors charged under the axial symmetry and $N_c$ the 
rank of the gauge group.  In general there are additional fermions that are uncharged
under the $U(1)_A$ symmetry.  

Before entering into the details of the calculations, it is important to characterize the physical meaning on the QFT side of the Lagrangian parameter space.

The {\bf bulk mass} $m^2$ determines the scaling dimension of the operator dual to $\Phi$, according to:
\begin{equation}
\Delta_\Phi= \frac{d+ \sqrt{d^2 + 4 m^2}}{2},
\end{equation}  
where $d=4$ for our case. Taking $m^2=-3$ gives $\Delta_\Phi= 3$, so that this operator has the dimension of a fermion bilinear massterm in four dimensions, and its source has dimension one, i.e. can be taken as a boundary mass $M$. If $\Delta_\Phi \neq 3$
we take appropriate powers of the non-normalizable mode to obtain a dimension one parameter
as in (\ref{dictionary}). 

The {\bf quartic coupling} $\lambda$ is a measure of the effective number of degrees of freedom which are not decoupled in the infrared in the trivial phase. This can be understood in terms of holographic relationship between the rank of the gauge group and the cosmological constant, as explained below. Generically in the trivial phase the scalar field runs 
to its value at the minimum of the potential $dV/d\phi=0$ modifying the effective IR cosmological constant. Since according to the holographic dictionary $L^4 \propto N$ 
with $N$ the total number of degrees of freedom

\begin{equation}
\left(\frac{N_{\rm IR}}{N_{\rm UV}}\right)_{\rm triv}= \frac{1}{\left(1+ \frac{( m^2)^2}{24 \lambda}\right)^2}
\end{equation}
At vanishing quartic coupling the theory loses all degrees of freedom in the IR and becomes
strongly coupled. For vanishing gravitational coupling the charged degrees of freedom
are negligible. 

% I filled the missing equations %
Finally, the {\bf charge} $q$ modulates the mixing between the operators dual to $\Phi$ and $A_M$. In the limit $q \to 0$ these two operators do not mix along the RG, and all the interesting physics is lost. This can be seen from the equations of motion for the axial field (\ref{eomaxial}). We denote from now on by $j^\mu$ and $j_5^\mu$ the r-dependent quantities which reduce to the consistent dual currents at the boundary: 
\begin{align}
j^\mu(r) &=\sqrt{-g} \left( F^{\mu r} + 4 \alpha \epsilon^{r \mu\nu\rho\sigma} A_\nu F_{\rho\sigma} \right), \label{smallj}\\ 
j^\mu_5(r) &=\sqrt{-g} \left( F_5^{\mu r} + \frac{4 \alpha}{3} \epsilon^{r \mu\nu\rho\sigma} A_\nu F_{\rho\sigma}^5 \right). \label{smallj5}
\end{align}
In terms of these quantities and dropping total derivatives the $\mu$ components of (\ref{eomaxial}) can be written as:
\begin{equation}
\label{mixing}
\frac{d}{dr}\left( j_5^\mu + \sqrt{-g}\frac{2}{3}\alpha  \epsilon^{r\mu\nu\rho\sigma}A_\nu^5 F_{\rho\sigma}^5\right)= 2\sqrt{-g}q^2 \phi^2 \left(A^\mu -\frac{1}{q}\partial^\mu \theta\right),
\end{equation}
where $\Phi=\phi e^{i \theta}$ and it is apparent how, in the limit $q \to 0$, all the properties of the axial coupling will match between the UV and the IR, as the left hand side of (\ref{mixing}) is a total radial derivative.

Explicit breaking of the axial $U(1)$ will be introduced by demanding that the
non-normalizable mode of the scalar does not vanish. In addition we introduce
a constant spatial component of the axial gauge field. 
Thus we use the boundary conditions 
\begin{equation}\label{dictionary}
\lim_{r \rightarrow \infty} r \Phi^\gamma = M, \hspace{0.5cm} \lim_{r \rightarrow \infty} A_z = b,
\end{equation}
where $\gamma^{-1} = 4 - \Delta_\Phi$ and $\Delta_\Phi$ is the dimension of the operator dual to $\Phi$. With these definitions  $M/b$ is a dimensionless parameter regardless of $\Delta_\Phi$.

We will be looking for solutions which are asymptotically AdS and present anisotropy in the $z$-direction because of the non-zero component of the background axial gauge field. We also expect an AdS black hole geometry for the finite temperature case. Thus, we will use the following ans{\"a}tze:
\begin{align}
T = 0 &:\hspace{0.5cm} ds^2 = u \left( -dt^2 + dx^2 + dy^2 \right) + \frac{dr^2}{u} + h dz^2, \hspace{0.5cm} \Phi = \phi , \hspace{0.5cm} A = A_z dz . \label{ansatzzero}\\
T \neq 0 &:\hspace{0.5cm} ds^2 = -u dt^2 + f \left( dx^2 + dy^2 \right) + \frac{dr^2}{u} + h dz^2, \hspace{0.3cm} \Phi = \phi , \hspace{0.5cm} A = A_z dz . \label{ansatzfinite}
\end{align}

When one tries to solve the equations of motion asymptotically near the horizon using this ansatz, one sees that the system has three kinds of solutions. These will turn out to correspond to the trivial phase, the non-trivial phase and the critical point located between them respectively. They will be labeled by the ratio $M/b$, which is  the only physically meaningful dimensionless parameter due to the conformal symmetry at zero temperature. 

\begin{figure}
\centering 
\resizebox{0.48\textwidth}{!}{\input{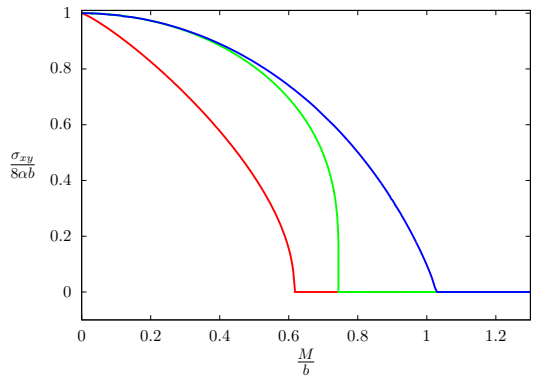}}
\caption{The zero temperature Hall conductivity for different values of the model's parameters: $m^2  =-2$, $\lambda=1/10$ (red), $m^2  =-3$, $\lambda=1/10$ (green), $m^2  =-3$, $\lambda=1$ (blue). It can be observed how the critical value for the $M/b$ parameter where the conductivity goes to zero changes in the different cases.}\label{fig:diffparams}
\end{figure}

In our study we will look for the value of $M/b$ corresponding to the critical solution as a function of the free parameters in the Lagrangian of the model ($m^2$, $q$ and $\lambda$). 
In general if the critical value is $(M/b)_c$ then for $(M/b) < (M/b)_c$ we will have a topologically nontrivial solution
as shown in figure \ref{fig:diffparams}.
When $M/b$ goes to infinity, it means that the trivial phase has shrunk to only one point.

\begin{figure}
\centering
\includegraphics[width=0.50 \textwidth]{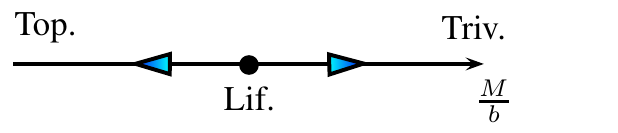}
\caption{Schematics of the RG flow of the model. At low energy an unstable critical point is present at a certain value of $M/b$, corresponding to a Lifshitz scaling solution. Small deviations in $M/b$ around this point will make the system flow in the infrared to either the topological phase to the left or to the trivial one to the right.}
\label{RGgood}
\end{figure}

\subsection{Universality}\label{sec:universality}

The first step to map ($m^2 $, $q$, $\lambda$) into the critical value $M/b$ is to solve the equations of motion asymptotically in the IR. Following \cite{Landsteiner:2015pdh} we
expect three solutions. Two of them will correspond to the two phases and the remaining one will correspond to the critical point that lies in the separation between them.

If one looks for an exact scaling solution in the IR, it is possible to obtain a Lifshitz-like geometry with a non-trivial scaling exponent in the $z$-direction, the {\bf critical solution}. In this section we will focus on this solution and on how we obtain $M/b$ for each choice of the parameters in the Lagrangian. 
In general there is only one critical solution for a given $m^2$, $\lambda$ and $q$.
For completeness we briefly explain how to obtain the other two solutions.
At the horizon either $A_z$ or $\phi$ must vanish, where $\phi=0$ holds for the topological phase and $A_z=0$
for the trivial one. 
Adding irrelevant perturbations leads to a two parameter space of solutions in the IR and by scaling symmetries
this gets reduced to one dimensionless parameter which in the UV determines $M/b$. 
The Hall conductivity is
\begin{equation}\label{eq:sigmaHall}
\sigma^H_{xy} = 8 \alpha A_z (0)\,.
\end{equation}
A solution with $A_z(0) \neq 0$ we call topological,  otherwise we call it trivial.

The {\bf critical solution} is obtained from a scaling ansatz on the fields: 
\begin{equation}
  u = u_0 r^{2 \alpha}, \hspace{0.5cm} h = h_0 r^{2 \beta}, \hspace{0.5cm} A_z = A_0 r^\gamma, \hspace{0.5cm} \phi = \phi_0 r^\delta. 
\end{equation}
Imposing equations of motion we find:
\begin{equation}
  u = u_0 r^2 , \hspace{0.5cm} A_z = r^\beta , \hspace{0.5cm} h = h_1 r^{2 \beta} , \hspace{0.5cm} \phi = \phi_0 , 
\end{equation}
where $h_1$ is $h_0/A_0^2$.

The four parameters ($u_0$, $h_1$, $\beta$ and $\phi_0$) are functions of ($m^2 $, $q$ and $\lambda$):
\begin{align}
0 &= 3 h_1 ( u_0 - 1 ) - \frac{1}{8} u_0 \beta^2 + \frac{1}{4} \phi_0^2 ( h_1 m^2 - q^2 ) + \frac{1}{8} \phi_0^4 h_1 \lambda , \\
0 &= 2 u_0 h_1 ( 1 - \beta ) - \frac{2}{3} q^2 \phi_0^2 , \\
0 &= 3 u_0 \beta - 2 q^2 \phi_0^2 , \\
0 &= m^2 h_1 + q^2 + \lambda h_1 \phi_0^2 .
\end{align}

Solving the last three equations for $u_0$, $h_1$ and $\beta$, we obtain the following relations:
\begin{align}
u_0 &= \frac{2 q^2 \phi_0^2}{3 \beta} \,, \\
h_1 &= - \frac{q^2}{m^2 + \lambda \phi_0^2} \,, \label{equationh1}\\
\beta &= - \frac{2 q^2}{m^2 + \lambda \phi_0^2 - 2 q^2} \,. \label{equationbeta}
\end{align}
The first equation turns into a third order equation in $\phi_0^2$.

It is necessary to impose physical constraints which can be summed up in the following way:
\begin{description}
 \item [Regularity:] $\beta$ needs to be larger than zero.
 \item [Reality:] $A_z$ and $\phi$ need to be real, because we want to recover a real value of $M$ and $b$ in the UV. This gives us the conditions that $\beta$ is real and $\phi_0^2$ is real and positive.
 \item [Null-energy condition:] We impose the null-energy condition $T^{MN} \xi_M \xi_N \geq 0$, with $\xi_M$ any future-pointing light-like vector field. Taking $\xi = \frac{1}{\sqrt{u}} dt + \sqrt{h} dz$ and making use of Einstein's equations gives $T^{MN} \xi_M \xi_N = G^{MN} \xi_M \xi_N = 1-\beta$, so the null-energy condition reads $\beta \leq 1$. This condition also enforces realness of the axial field since it ensures the positivity of $h_1$, as it becomes obvious from (\ref{equationh1}) and (\ref{equationbeta}) that $\beta = 2 / ( 2 + h_1^{-1} )$. 
\end{description}

Once the physical solution of the system has been found, one can flow to asymptotic AdS in the UV by perturbing the system with irrelevant perturbations around the Lifshitz fixed point:
\begin{equation}
 \begin{aligned}
  u &= u_0 r^2 ( 1 + \delta u\, r^\chi ), \\
  h &= h_1 r^{2\beta} ( 1 + \delta h\, r^\chi ), \\
  A_z &= r^\beta ( 1 + \delta a\, r^\chi ), \\
  \phi &= \phi_0 ( 1 + \delta \phi\, r^\chi ). \\
 \end{aligned}
\end{equation}
The fact that all the perturbations have the same scaling exponent $\chi$ follows immediately from the linearized form of the equations of motion. Solving these again for the new parameters ($\chi$, $\delta u$, $\delta h$, $\delta a$, $\delta \phi$) it is important to enforce realness and regularity on this solutions and this requires $\chi$ to be real and positive. The equation has seven solutions but there is only one solution satisfying these two conditions. The other four parameters can be expressed as a function of only one of them. The possibility to integrate numerically the solution to the UV using the equations of motion will only depend on the sign of this free parameter. Finally, the boundary conditions (\ref{dictionary}) allows us to complete the map from ($m^2 $, $q$, $\lambda$) to $M/b$ for the critical solution and study the appearance and location of the phase transition.

Aside from this computation we also present a more physical argument that comes directly from the holographic dictionary in order to justify our results.

In the holographic setup we can read the number of degrees of freedom directly from the gravitational geometry. The starting point is the identification:
\begin{equation}\label{dof}
N= \left( \frac{L}{l_s}\right)^4,
\end{equation}
where $L$ is the Anti de Sitter length and $l_s$ is the string length.

The UV/IR correspondence of holography maps the geometry near the boundary of spacetime to the ultraviolet conformal fixed point of the dual QFT, while the deep bulk geometry (which in the finite temperature case is a black hole's horizon) contains information about the infrared degrees of freedom. If both the ultraviolet and infrared geometries are asymptotically AdS then (\ref{dof}) implies:
\begin{equation}
\frac{N_{\rm IR}}{N_{\rm UV}}= \left(\frac{L_{\rm IR}}{L_{\rm UV}}\right)^4,
\end{equation}
where $N_{\rm UV/\rm IR}$ denotes the UV/IR degrees of freedom and $L_{\rm UV/\rm IR}$ the UV
and IR Anti de Sitter length scales. Of course, in order for classical gravity to be a valid description, both $N_{\rm UV}$ and $N_{\rm IR}$ are formally infinite, but their ratio remains a finite quantity.
$L_{\rm IR}$ is defined implicitly through the infrared cosmological constant $\Lambda_{\rm IR}= -\frac{12}{L_{\rm IR}^2}$ whose value can be computed explicitly from $\Lambda_{\rm IR}= -\frac{12}{L_{\rm UV}^2}+ V(\Phi_{\rm IR})$. The counting of degrees of freedom can be expressed as:
\begin{equation}\label{dofLambda}
\frac{N_{\rm IR}}{N_{\rm UV}}= \left( \frac{\Lambda_{\rm UV}}{\Lambda_{\rm IR}}\right)^2,
\end{equation}
or better, in terms of the scalar curvature (since AdS is an Einstein spacetime):
\begin{equation}\label{dofR}
\frac{N_{\rm IR}}{N_{\rm UV}}= \left( \frac{R_{\rm UV}}{R_{\rm IR}}\right)^2.
\end{equation}
Notice that this quantity can never be greater than one as we flow between two Anti de Sitter spacetimes, since it would violate 
the holographic version of the celebrated C-theorem\footnote{In the cases cited the gravitational metric, in the Fefferman-Graham gauge, is of the form $ds^2 = e^A(r)dx^\mu dx_\mu +dr^2$. The C-theorem states simply that $A''(r)\geq 0$, so that $A'(r)$ is an increasing quantity. In these geometries the scalar curvature is $R=-5 A'(r)^2 -4 A''(r)$ and if the space reaches an AdS fixed point at $r=r_i$, then $A'(r_i)=\frac{2}{L_i}$, $A''(r_i)=0$. At these points $R(r_i)=-20/L_i^2$ and the growing behavior of $A'(r)$ implies  the increase of the modulus  of the scalar curvature as we flow away from the conformal boundary.} \cite{Girardello:1998pd, Freedman:1999gp}.

While this reasoning applies in a straightforward way in the case of AdS asymptotics, it has to be generalized if the IR geometry is not AdS, as is the case for our critical solution. 
On physical grounds we use the scalar curvature as an estimator of the degrees of freedom, as it is the only constant length 
scale which is physically observable. 

In our holographic model there are three different possible infrared geometries which all have different curvatures. Using (\ref{dofLambda}) and (\ref{dofR}), a straightforward computation gives:
\begin{align}
\left( \frac{N_{\rm IR}}{N_{\rm UV}} \right)_{\rm top} &= 1, \\
\left( \frac{N_{\rm IR}}{N_{\rm UV}} \right)_{\rm triv} &= \frac{1}{\left(1+ \frac{(m^2)^2}{24 \lambda}\right)^2}, \\
\left( \frac{N_{\rm IR}}{N_{\rm UV}} \right)_{\rm c} &= \left(\frac{10}{u_0(6 +\beta(3+\beta))}\right)^2 \label{eq:Ncrit}\,.
\end{align}
Notice that only the result for the critical solution depends on the charge $q$.
Consistency with our weak coupling intuition of the gradual ``gapping'' of the system then demands:
\begin{equation}
\left( \frac{N_{\rm IR}}{N_{\rm UV}} \right)_{\rm top}> \left( \frac{N_{\rm IR}}{N_{\rm UV}} \right)_{\rm c}> \left( \frac{N_{\rm IR}}{N_{\rm UV}} \right)_{\rm triv}\,.
\end{equation}

\begin{figure}
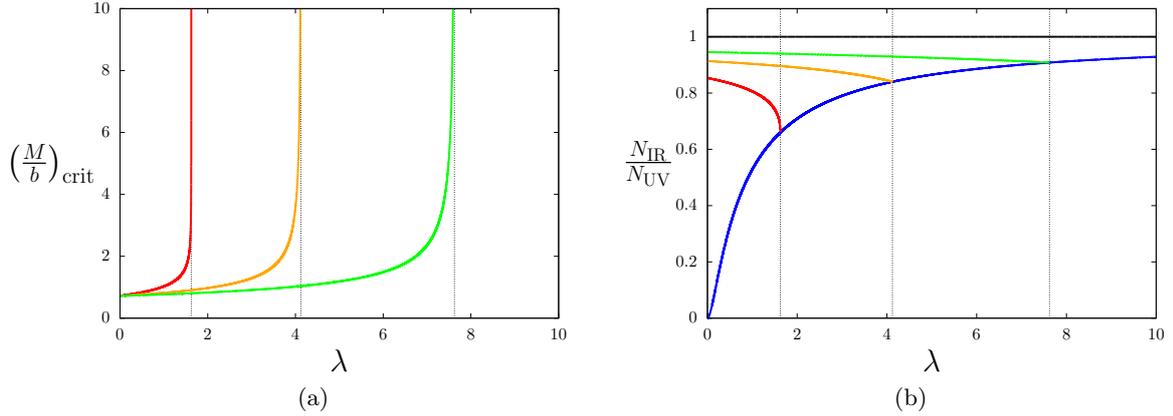

\centering
\subfloat[]{\resizebox{0.48\textwidth}{!}{\input{lambda-map1.tex}}}\hfill \subfloat[]{\resizebox{0.48\textwidth}{!}{\input{lambda-dof1.tex}}}
\caption{Phase transition as a function of the model's parameters: (a) critical $M/b$ as a function of $\lambda$, (b) infrared degrees of freedom as a function of $\lambda$.  Divergences in the $M/b$ values and crossing in the counting of degrees of freedom signal the impossibility for the phase transition to take place. They are emphasized by a dotted vertical line and coincide between the two methods. Various values of the charge $q\geq 1$ are plotted: $q=1$ (red), $q= 1.5$ (orange) and $q=2$ (green).}
\label{plotlambda}
\end{figure}

\begin{figure}
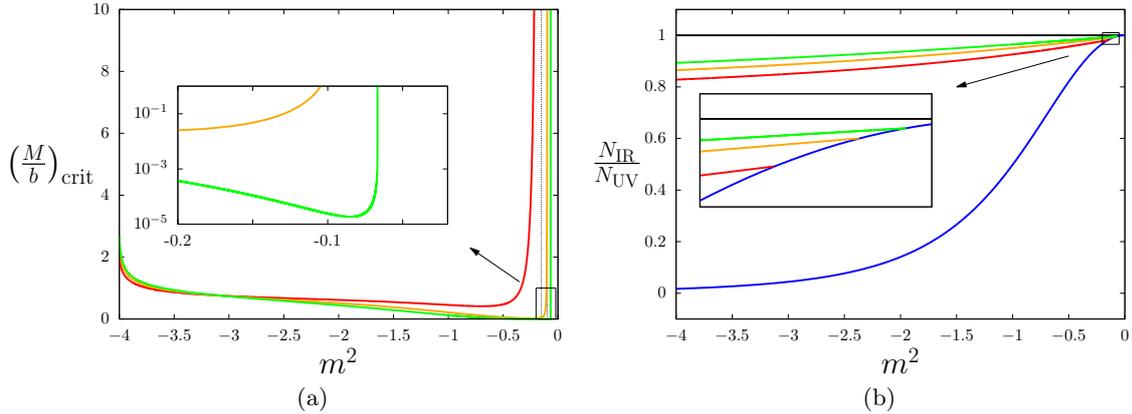

\centering
\subfloat[]{\resizebox{0.48\textwidth}{!}{\input{mass-map1.tex}}} \hfil 
\subfloat[]{\resizebox{0.48\textwidth}{!}{\input{mass-dof1.tex}}}
\caption{Phase transition as a function of the model's parameters: (a) critical dimensionless ratio $M^\gamma/b$ ($\gamma^{-1}=4-\Delta_{\Phi}$) as a function of $m^2 $, (b) infrared degrees of freedom as a function of $m^2$. Points at which the transition ceases to take place are signaled by a dotted vertical line and coincide between the two models. Various values of the charge $q\geq 1$ are plotted: $q=1$ (red), $q= 1.25$ (orange) and $q=1.5$ (green).}
\label{plotmass}
\end{figure}

The result of the two methods described above are presented in figures \ref{plotlambda} and \ref{plotmass}, where we study dependence on $\lambda$ and $m^2$, respectively.
 
In the case of the dependence on $\lambda$ we clearly see that, for large enough $\lambda$, the critical $M/b$ parameter diverges. This implies, as the topological phase still exists for lower $M/b$ than the critical one, that the transition to the trivial semimetal phase ceases to take place, making the system impossible to be gapped at low energies. It is interesting that the limit $\lambda \to 0$ reaches smoothly a finite constant value of $M/b$. However, this is not easy to analyze, as the gravitational backreaction on the geometry has to be arbitrarily big in order to compensate for the scalar field's energy density, which scales as $\lambda^{-1}$. 

The argument using the infrared degrees of freedom strengthens our conclusions. In fact, it can be seen that, as we increase $\lambda$, the infrared degrees of freedom of the critical solution become eventually less than those of the trivial one. In this case the phase transition ceases to take place, as the gapped phase would have too many degrees of freedom to be reached from the critical point and we would be left with a system ``always'' in the topological phase. Furthermore, the limiting $\lambda$ values coincide between the two computations within numerical precision.
 
Looking now at figure \ref{plotmass}, we notice two things. First of all, we get a similar steep curve to the one in the left panel of figure \ref{plotlambda} when we approach a critical lower value for $m^2$. This again signals the vanishing of the trivial phase, as the mass operator crosses the marginality bound. This crossing is forbidden in the strict $m^2 =0$ limit. In this case the axial current acquires an anomalous dimension in the UV changing the RG flow drastically\footnote{This is the holographic St\"{u}ckelberg mechanism whose interpretation is that the anomaly
has contributions from dynamical gauge fields \cite{Klebanov:2002gr,Casero:2007ae,Gursoy:2014ela,Jimenez-Alba:2014iia}. The dimensionless coupling $M=\lim_{r \to + \infty} \phi(r)$ acts as an effective mass for the gauge field $A_M$ through the quadratic gauge coupling, giving $m_A= q^2 M^2$.}. Notice from the inset in figure \ref{plotmass} (a) that the lower bound $(M/b)_c >0$ does not get crossed in any case, so that we never find a vanishing topological phase.

This is confirmed by the counting of degrees of freedom, which crosses the ones from the trivial phase at the same value of $m^2 $. If, on the other hand we had lost the topological phase, we would have seen a crossing of the horizontal line $N_{\rm IR}/N_{\rm UV}=1$. The limit $m^2 \to -4$ does not present any particular problem\footnote{This conclusion is however not complete, as we describe only dual operators with dimensions $2\leq \Delta_\Phi \leq 4 $. Operators of dimensions $2\leq \Delta_\Phi \leq 1$ can be described using the alternative quantization scheme for big enough negative masses ($-4 \leq m^2  \leq -3$ for a scalar field in five dimensions). In this case we would identify the leading contribution at the conformal boundary as a fixed one point function $\langle \mathcal{O}_\Phi \rangle$ for the operator dual to $\Phi$ rather than its fixed coupling.}.
 
 As a final remark we notice that the system seems to display an odd behavior in the case $q^2<1$, for which two critical solutions with different infrared degrees of freedom can be found. These flow to two separate values of $M/b$ for which the system admits no solution in between (as can be seen either numerically or by counting degrees of freedom). This unphysical situation may be the consequence of some fundamental bound on gravitational systems, with profound consequences on the allowed dual operator's charges. We leave this problem for future studies.

\section{Axial conductivity and infrared screening}\label{sec:sigmaaxial}

The holographic model of section \ref{sec:qpt} has a nonzero transversal conductivity in the topologically nontrivial phase. 
\begin{equation}\label{AHE}
\mathcal{J}^i= \epsilon^{ijk}\sigma^H_j E_k \, ,
\end{equation}
where $\sigma^H_i$ is the electric anomalous Hall conductivity. 

From a quantum field theoretical point this is a consequence of the anomaly giving rise to a Chern-Simons like coupling:
\begin{equation}\label{effectiveaction}
W[V,A,b]=W[V,A,0] + \frac{N_A^f N_c^2}{24 \pi^2}\int d^4 x \epsilon^{\mu\nu\rho\sigma} b_\mu \left( 3 V_\nu F_{\rho\sigma} + A_\nu F_{\rho\sigma}^5\right) \,,
\end{equation}
where the parameter $b_\mu$ is the effective infrared value of the axial gauge field \cite{Grushin:2012mt}.
In holography this result can be extracted in a rather straightforward way from the five dimensional Maxwell-Chern-Simons equations of motion (\ref{eomvector}). When expressed in terms of the quantity $j^\mu$ from (\ref{smallj}), the $\mu$ components of (\ref{eomvector}) read:
\begin{equation}
\frac{d}{dr}j^\mu +\sqrt{-g}\nabla_\nu F^{\nu\mu} + 8 \alpha \sqrt{-g}\epsilon^{\mu\nu\rho r \sigma }\partial_\nu \left(A_\rho F_{r\sigma} \right) =0. \label{noname}
\end{equation}
We now turn on a boundary electric field which has a finite zero frequency and momentum value. Performing a Fourier transform in the Minkwoski coordinates and taking the explicit $\omega, \ \vec{k} \to 0$ limit, (\ref{noname}) gives the conservation of the current zero mode $\tilde{j}^\mu= \int d^4 x j^\mu$:
\begin{equation}
\frac{d}{dr}\tilde{j}^\mu= 0. 
\end{equation}
Upon radial integration between the boundary and the horizon one gets, taking into account (\ref{vectorcurrent}):
\begin{equation}
\tilde{\mathcal{J}}^\mu= \tilde{j}^\mu (r_H).
\end{equation}
This conservation equation allows us, in the zero frequency limit, to evaluate the vector current directly in terms of horizon quantities. According to the holographic renormalization group, the latter describe the low energy degrees of freedom of the theory. 

The explicit dependence of the electric field on the horizon from the one at the boundary can be recovered, at zero frequency and momentum, from the Fourier transformed Bianchi identity $\epsilon^{LMNPQ} \partial_N F_{PQ}=0$, taking $L,M$ to be two of the spatial directions:
\begin{equation}
 \frac{d}{dr}\tilde{F}_{0i}= \frac{d}{dr}\tilde{E}_{i}=0.
\end{equation}
When evaluating $j^\mu$ at the horizon, the $F^{r\mu}$ term can be expressed as a function of the electric field through the infalling boundary conditions following \cite{Iqbal:2008by}. This term gives rise to a longitudinal conducitivity, which vanishes at zero temperature and will be omitted for simplicity.
Finally, explicit evaluation of the Chern-Simons part of (\ref{vectorcurrent}) on the horizon gives, for the transversal part:
\begin{equation}
\mathcal{\tilde{J}}^i= 8\alpha \epsilon^{ijk}A_j(r_H) \tilde{E}_k,
\end{equation}
which, upon fixing $\alpha=\frac{N_f^A N_c^2}{16\pi^2}$ and identifying $A_j(r_H)$ with the effective infrared coupling $b_i$, is in perfect agreement with (\ref{AHE}).  

In this section we focus on computing the Hall conductivity for the anomalous axial current $\mathcal{J}^\mu_5$ and the calculation in this case has to be done resorting to numerical methods, because of the coupling to the scalar field $\phi$. However, as we will see towards the end of the section, we would expect the structure of the anomalies to completely determine the form of such a coefficient because of the relation (\ref{effectiveaction}). This is, in fact, possible. Properly taking into account the RG flow to low energies allows us to successfully match the holographic computation to the weak coupling results. 

These axial transport properties are of high physical interest, as they can appear in condensed matter studies. Axial vector fields arise at an effective level due to lattice strain\cite{Cortijo:2016wnf,Pikulin:2016wfj,axialmagneticAdolfo}. This computation also describes a topological signature of anomalous $U(1)^3$ theories in the presence of charged matter. This can then be used to directly detect the family of phase transitions presented in section \ref{sec:qpt} even if only the anomalous symmetry is present. Applications of this idea will be presented in the last section in which we explicitly analyze a consistent string-theoretical realization.

\subsection{Axial Hall conductivity}

The axial Hall conductivity cannot be computed easily by solving the RG flow of the current because of the symmetry breaking
of the scalar field. We therefore resort to the usual method of computing the conductivities via Kubo formulae\footnote{In what follows $\sigma_{jk}$ is related to the $\sigma^H_i$ conductivity through $\sigma^H_i=\frac{1}{2} {\epsilon_i}^{jk}\sigma_{jk}$.}:
\begin{align}
 \sigma_{ik} &= \lim_{\omega \to 0} \frac{1}{i\omega} \langle \mathcal{J}_i \mathcal{J}_k \rangle \ (\omega, \vec{k} = 0 ),\\
 \sigma_{ik}^5 &= \lim_{\omega \to 0} \frac{1}{i\omega} \langle \mathcal{J}_i^5 \mathcal{J}_k^5 \rangle \ (\omega, \vec{k} = 0 ) \label{kubo}.
\end{align}

It has been shown that in holography one can obtain the retarded Green's function by studying the fluctuations around the background for the gauge fields dual to the currents and imposing infalling boundary conditions. In this section we are interested only in the axial conductivity but for completeness reasons it is interesting to also include the equations of motion for the vector perturbations.

The Hall conductivity is the off-diagonal part of the above equation so, for the vector conductivity, we need to turn on fluctuations on the $x$- and $y$-directions of the vector gauge field i.e. $\delta V_{x/y} = v_{x/y} (r) e^{-i \omega t}$. The equations of motion for these fluctuations are both for finite and zero temperature:
\begin{equation}
 \begin{aligned}
  v_x'' + \left( \frac{h'}{2 h} + \frac{u'}{u} \right) v_x' + \frac{\omega^2}{u^2} v_x + \frac{8 i \omega \alpha Az'}{u \sqrt{h}} v_y & = 0 , \\
  v_y'' + \left( \frac{h'}{2 h} + \frac{u'}{u} \right) v_y' + \frac{\omega^2}{u^2} v_y - \frac{8 i \omega \alpha Az'}{u \sqrt{h}} v_x & = 0 . 
 \end{aligned}
\end{equation}

However, the axial gauge field fluctuations ($\delta A_{x/y} = a_{x/y} (r) e^{-i \omega t}$) will also produce fluctuations on the metric ($\delta g_{xz} = g_{xx}\ {h^x}_z e^{-i \omega t}$ and $\delta g_{yz} = g_{yy}\ {h^y}_z e^{-i \omega t}$) and the complete set of equations of motion will have the following form:
\begin{equation}
 \begin{aligned}
 a_x'' + \left( \frac{h'}{2 h} + \frac{u'}{u} \right) a_x' + \frac{\omega^2}{u^2} a_x + \frac{8 i \omega \alpha A_z'}{u \sqrt{h}} a_y - \frac{2 q^2 \phi^2}{u} a_x - \frac{f A_z'}{h} {h^x}_z' &= 0 , \\ 
 a_y'' + \left( \frac{h'}{2 h} + \frac{u'}{u} \right) a_y' + \frac{\omega^2}{u^2} a_y - \frac{8 i \omega \alpha A_z'}{u \sqrt{h}} a_x - \frac{2 q^2 \phi^2}{u} a_y - \frac{f A_z'}{h} {h^y}_z' &= 0 , \\ 
 {h^x}_z'' + \left( \frac{2 f'}{f} + \frac{u'}{u} - \frac{h'}{2 h} \right) {h^x}_z' + \left( \frac{f''}{f} - \frac{u''}{u} + \frac{f' h'}{2 f h} - \frac{u' h'}{2 u h} \right) {h^x}_z + \frac{\omega^2}{u^2} {h^x}_z + \frac{A_z'}{f} a_x' + \frac{2 q^2 A_z \phi^2}{f u} a_x &= 0 , \\
 {h^y}_z'' + \left( \frac{2 f'}{f} + \frac{u'}{u} - \frac{h'}{2 h} \right) {h^y}_z' + \left( \frac{f''}{f} - \frac{u''}{u} + \frac{f' h'}{2 f h} - \frac{u' h'}{2 u h} \right) {h^y}_z + \frac{\omega^2}{u^2} {h^y}_z + \frac{A_z'}{f} a_y' + \frac{2 q^2 A_z \phi^2}{f u} a_y &= 0 , \\
 \end{aligned}
\end{equation}
from which the zero temperature case follows from taking $f$ to be equal to $u$. Besides the coupling to the gravity sector, the main difference between both sets of equations is that in the axial gauge field case we cannot take the equation of motion to be a total derivative, in the same way it already happened to the background field $A_z$ (see equation (\ref{mixing})). On contrary, a term involving the scalar field appears, giving a non-trivial RG flow along the bulk which will take a major role in the understanding of the results obtained for the conductivity.

The coupling to the gravity sector cannot be avoided. We can make both equations for the vector field decouple from each other with a particular set of basis i.e. $v_\pm = v_x \pm i v_y$. However, with an analogous basis one can only reduce the set of four equations of the axial field to two sets of two equations each. Therefore, the linear response coefficients relating the non-normalizable mode with the normalizable mode will have to be obtained taking care of holographic operator mixing \cite{Kaminski:2009dh}. We will try to briefly explain the method followed. From now on, we focus on the axial gauge field sector.

We start by expanding the action to second order in perturbations of the fields. Then, we Fourier transform this quantity and take only positive momenta. 

At this stage we multiply the fields by the proper power of $r$, to make their leading term in the near-boundary asymptotics constant. Finally, having the action expressed in this way, we impose the equations of motion and get the on-shell action, which will have the following form:
\begin{equation}
 S = \int dk_> \left[ 2 A_{IJ} \Phi_{-k}^I {\Phi'}_{-k}^J + B_{IJ} \Phi_{-k}^I \Phi_{k}^J \right]_{r_h}^{r_b} = \int dk_> \left[ \varphi_{-k}^I {\cal F}_{IJ}(k,r) \varphi_k^J \right]_{r_h}^{r_b}.
\end{equation}
where $\Phi_k^I$ is the field mode associated to momentum $k$ and $\varphi_k^I$ is its value at a cut-off close to the boundary that we will call $r_{\Lambda}$. The remaining key ingredient is to use the bulk-to-boundary propagator (BBP) to connect the first form of the on-shell action with the second one, by expressing the bulk fields $\Phi_k^I(r)$ in terms of their value at the cut-off $\varphi_k^I$:
\begin{equation}
\begin{aligned}
 \Phi_k^I (r) &= {F^I}_J (k,r) \varphi_k^J , \\
 \Phi_{-k}^I (r) &= {F^I}_J (-k,r) \varphi_{-k}^J = \varphi_{-k}^J {F^\dagger_J}^I (k,r).
\end{aligned}
\end{equation}

With all this, we can then take
\begin{equation}
 {\cal F} (k,r) = 2 F^\dagger A F' + F^\dagger B F,
\end{equation}
which will give us minus the retarded Green's function in the limit where the cut-off is shifted to the boundary:
\begin{equation}
 G_{IJ}^R (k) = - \lim_{r \to \infty} {\cal F}_{IJ} (k,r).
\end{equation}

The remaining piece of the method is the construction of the BBP. When one solves asymptotically the equations of motion near the horizon, it can be seen that each field is defined up to an unspecified integration constant. We can take these values for the different fields to form a vector and use a basis of this vector space. The normalization of this basis doesn't matter because we will end up normalizing the BBP to be the unit matrix at the cut-off. Therefore, for simplicity we will take the linearly independent combinations of the integration constants to be sets of ones with a minus one that is in a different position for each solution.

For each of the elements of this basis, we will integrate numerically the fields to the UV and construct a solution matrix $H(k,r)$ where the rows represent each fluctuation and the columns represent each of the different solutions:
\begin{equation}
 {H^I}_J (k,r) = {\Phi^I}_{(J)} (k,z).
\end{equation}
Finally, the BBP can be obtained as:
\begin{equation}
 F (k,r) = H(k,r) \cdot H(k,r_\Lambda)^{-1}.
\end{equation}

Now we have explained the method, we will include the field vector and the $A$ and $B$ matrices in the second-order on-shell action for finite temperature. Zero temperature case would be obtained substituting $f$ by $u$. In our case, we don't need to take into account the power of $r$ on the leading term of the asymptotic expansion, since $a_{x/y}$ have a constant leading term and, although the leading term of the metric perturbations will have a $r^2$ leading term, the variables ${h^x}_z$ and ${h^y}_z$ also have a constant leading term. 
\begin{equation}\label{matrices}
\begin{aligned}
\Phi &= \left(
 \begin{array}{c}
  a_x \\
  {h^x}_z \\
  a_y \\
  {h^y}_z \\
 \end{array}
  \right), \\\\
 A &= \left( 
 \begin{array}{c c c c}
  -\frac{\sqrt{h} f}{2} & 0 & 0 & 0 \\
  0 & - \frac{f^3}{4 \sqrt{h}} & 0 & 0 \\
  0 & 0 & -\frac{\sqrt{h} f}{2} & 0 \\
  0 & 0 & 0 & - \frac{f^3}{4 \sqrt{h}} \\
 \end{array}
 \right),\\\\
 B &= \left( 
 \begin{array}{c c c c}
  0 & \frac{f^2 A_z'}{\sqrt{h}} & - \frac{8 i \omega \alpha A_z}{3} & 0 \\
  0 & - \frac{3 f^2 f'}{2 \sqrt{h}} & 0 & 0 \\
  \frac{8 i \omega \alpha A_z}{3} & 0 & 0 & \frac{f^2 A_z'}{\sqrt{h}} \\
  0 & 0 & 0 & - \frac{3 f^2 f'}{2 \sqrt{h}} \\
 \end{array}
 \right).
\end{aligned}
\end{equation}

The application of this methodology requires the construction of a solution for the background fields and the study on top of them of solutions to the equations of motion of the fluctuations. The background fields are obtained following the method explained in previous section, where the IR solutions of both phases and of the critical point are integrated numerically to the UV. Then, we look for the asymptotic solutions of the fluctuations near the horizon and impose regularity and infalling boundary conditions, which could alternatively be seen as regularity on Eddington-Finkelstein coordinates. Thus, we have an analytical form for the boundary conditions that we will use for the numerical integration of the equations of motion of the fluctuations. It depends only on the holographic coordinate $r$, on the frequency $\omega$ and on the integration constants to which we made reference above, which will be either ones or minus ones. The numerical solutions of the fluctuations are then plugged into the $H$ matrices, the background fields are included in the $A$ and $B$ matrices and all this allows us to get the Green's function.

We can safely take $b$ to be the relevant physical energy scale throughout our computation. Therefore the only dimensionless parameter in the problem is $M/b$ for zero temperature and both $M/b$ and $T/b$ for the finite temperature case, as well as the frequency $\omega/b$ for the perturbations. When taking the small frequency limit we need to take care of the region in which we will integrate our equations of motion. In particular, we need to take the frequency to be smaller than the distance $r$ at which we took the IR limit of the integration region. The system is also very sensitive to the maximum $r$ at which we integrate, since the metric fluctuations give notable numerical errors for large integration regions. Therefore, since the solutions we obtain for these fields have the form of a domain wall and the final value is already reached at a quite low value of $r$ we looked at the value of the holographic coordinate at which the boundary value of the fluctuations was saturated and take this as our $UV$ result.

\subsection{Current renormalization.}
Before analyzing the results themselves it is useful to offer some weak coupling predictions for the outcome. As shown in \cite{Goswami:2012db,Zyuzin:2012tv,Grushin:2012mt}, the anomalous Hall effect stems from an anomalous contribution to the low energy effective action given by the infrared coupling to the chiral current $b^{\rm IR}_\mu J^\mu_5$\footnote{In this section we use the suffix IR to denote couplings of the effective low energy description of the system, in order to distinguish them from the ``bare'' couplings in the UV description.}. This current insertion can be ``traded'' as in (\ref{effectiveaction}) for the extra contribution to the effective action:
\begin{equation}
\delta W[V,A]= \frac{N^A_f N_c^2}{24 \pi^2}\int d^4 x \epsilon^{\mu\nu\rho\sigma} b^{\rm IR}_\mu \left( 3 V_\nu F_{\rho\sigma} + A_\nu F_{\rho\sigma}^5\right).
\end{equation}
Functional differentiation then gives a closed expression for both the anomalous axial and vector conductivities in the presence of an (axial) electric field $E^i_{(5)}$:
\begin{align}
\vec{\mathcal{J}} &= \frac{N^A_f N_c^2}{2 \pi^2}\ \vec{b}_{\rm IR} \times \vec{E},\\
\vec{\mathcal{J}}_5 &= \frac{N^A_f N_c^2}{6 \pi^2}\ \vec{b}_{\rm IR} \times \vec{E}_5,
\end{align}
from which we can read the transverse axial and vector transverse conductivities:
\begin{equation}
 \sigma_A= \frac{N^A_f N_c^2}{6 \pi^2}b_{\rm IR}, \hspace{0.5cm} \sigma_V= \frac{N^A_f N_c^2}{2\pi^2}b_{\rm IR}.
 \end{equation} 
This allows a prediction of the ratio between the axial transverse conductivity $\sigma_A$ and the vector one $\sigma_V$:
\begin{equation}
\frac{\sigma_A}{\sigma_V}=\frac{1}{3}.
\end{equation}
This ratio should be fixed completely by the structure of the anomalies of the theory and not receive any further low energy corrections, even though the low energy effective action is hard to determine precisely as the system is gapless. 

\begin{figure}
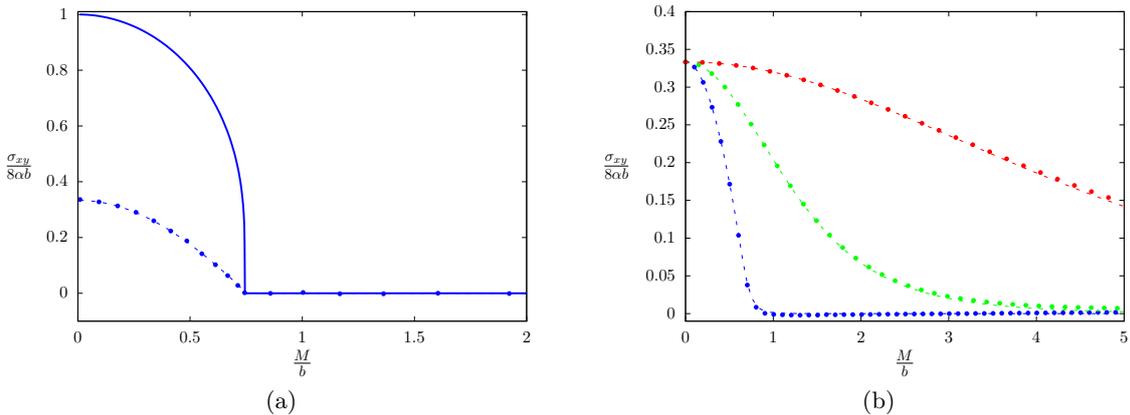

\subfloat[]{\resizebox{0.48\textwidth}{!}{\input{conductivities-zerotemperature1.tex}}}\hfill \subfloat[]{\resizebox{0.48\textwidth}{!}{\input{conductivities-finitetemperature1.tex}}}
\caption{Normalized transversal axial conductivity for both zero (a) and finite (b) temperatures. The line denotes the original result for the vector conductivity, dots denote numerical data for the axial conductivity and dashed lines denote the prediction (\ref{prediction}). The computation of (b) is done for different temperatures to prove the state independence of the result. In particular, here it is presented for $T/b=0.05$ (blue), $T/b=0.5$ (green), $T/b=2$ (red).}
\label{cond}
\end{figure}

Nevertheless, if we naively computed this quantity from holography, we would find a rather different answer (see figure \ref{cond}). To understand why this happens, let us remember that the weak coupling reasoning was based on a low energy effective field theory. As such, the external fields in the infrared will in general couple with a different strength than they do in the ultraviolet and their infrared coupling will also appear in the DC conductivities, as they are a response of the low energy physics.

\begin{figure}

\subfloat[]{\includegraphics[width=0.48\textwidth]{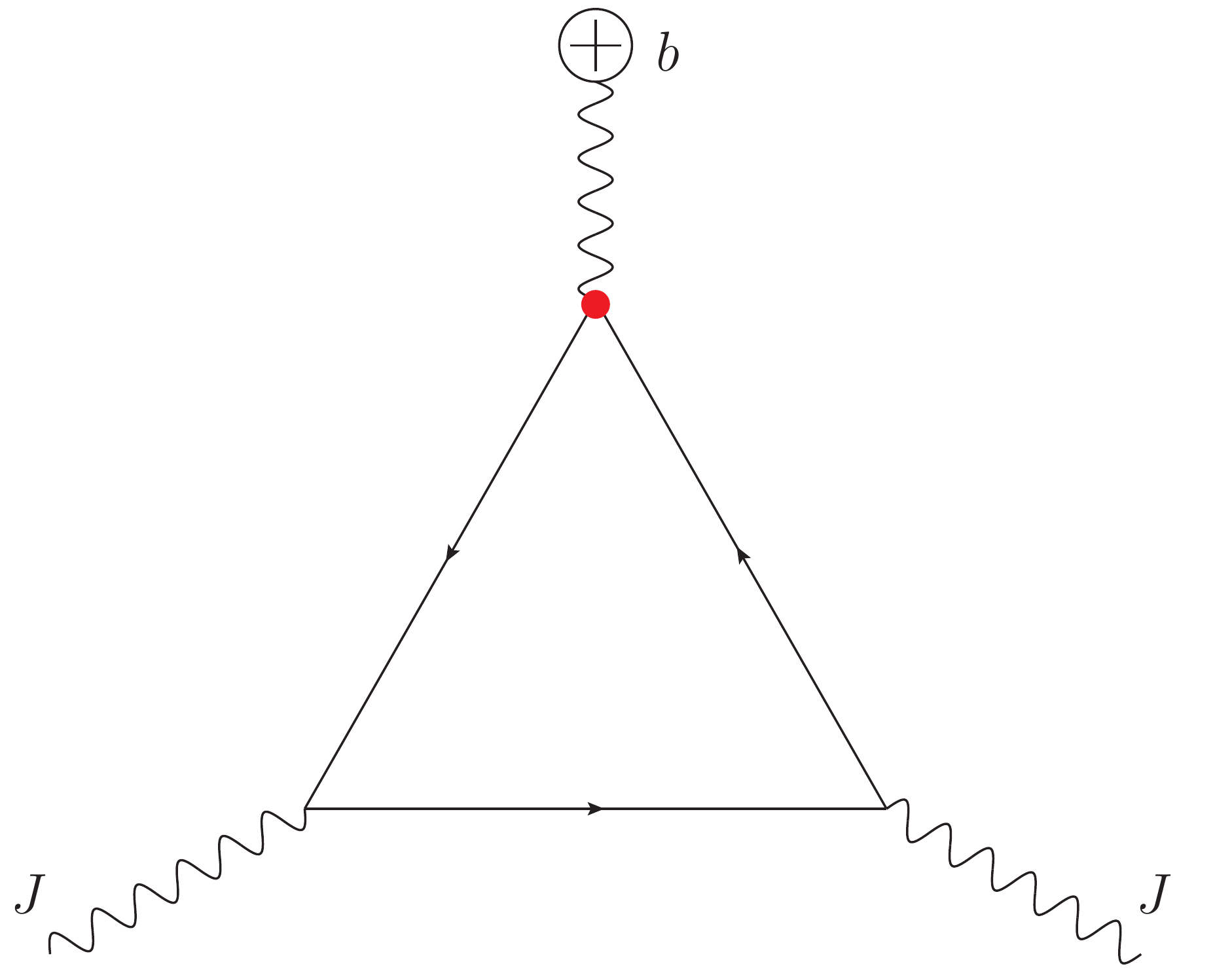}} \hfill \subfloat[]{\includegraphics[width=0.48\textwidth]{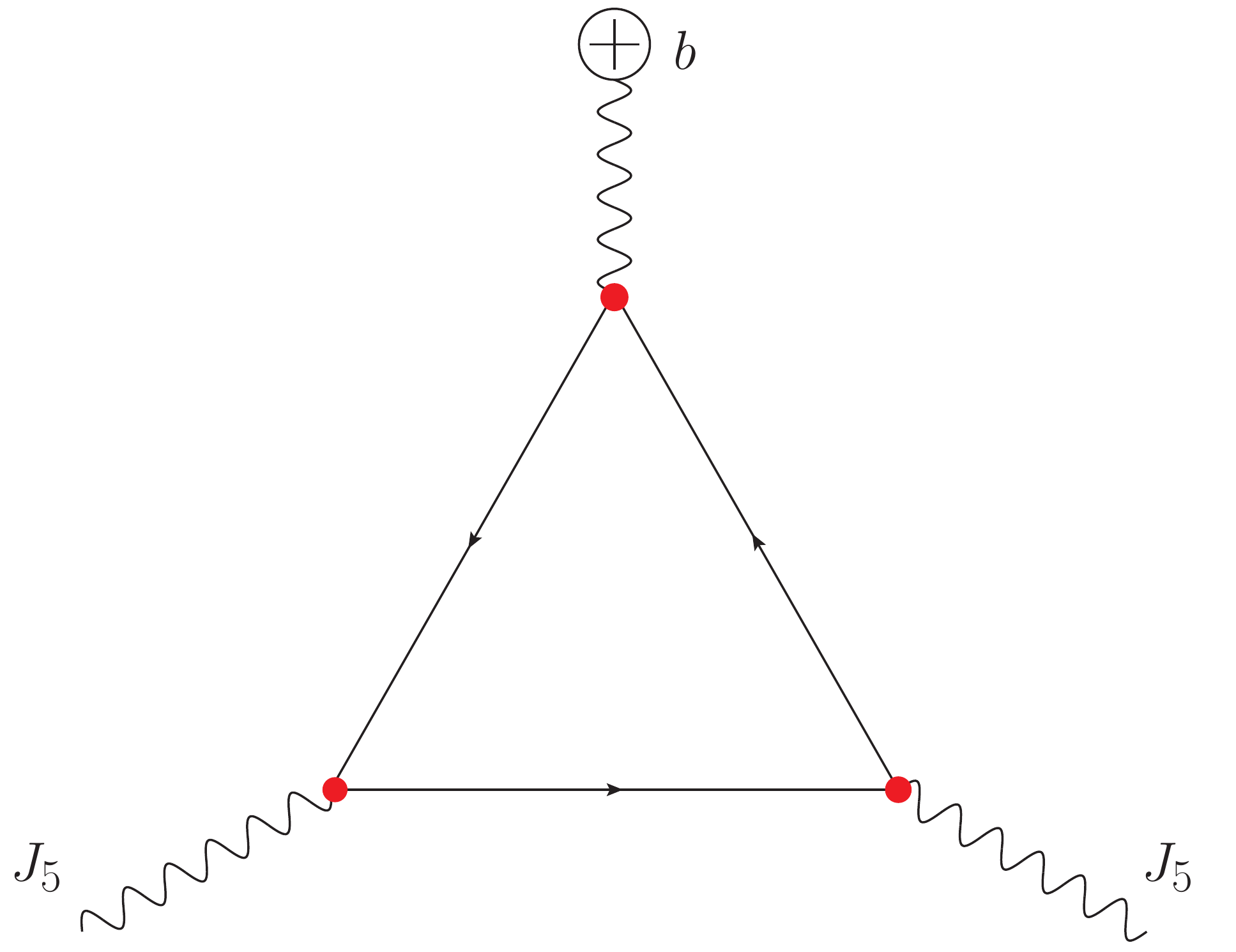}}
\caption{Diagrams corresponding to the leading corrections to the two-point functions $\langle J J \rangle$ and $\langle J_5 J_5 \rangle$. These will be the most important contributions to vector (a) and axial (b) conductivities in the infrared effective theory. Red dots denote the renormalized coupling of the axial operators $\sim \sqrt{Z_A}$. It can be noted how the first diagram scales as $\sqrt{Z_A}$, while the other one scales as ${\sqrt{Z_A^3}}$.} 
\label{diagrams}
\end{figure}

This can be thought of as a field renormalization effect due to the fact that we have sourced a charged operator in the UV theory (which is described by the scalar field $\Phi$). It is useful to introduce a renormalization constant $\sqrt{Z_A}$ such that:
\begin{equation}
\sqrt{Z_A} b_\mu= b^{\rm IR}_\mu.
\end{equation}

A complementary way of characterizing this phenomenon is to take into account the coupling  renormalization as we flow to the IR effective theory in the Wilsonian sense. In this case the important contributions come from the diagrams in figure \ref{diagrams}, where red dots account for the renormalized couplings.

Thus we will take all the axial fields to be renormalized through $\sqrt{Z_A}$ in the infrared, so that the effective action reads:
\begin{equation}
\delta W[V,A]= \frac{N^A_f N_c^2}{24 \pi^2} \int d^4 x \epsilon^{\mu\nu\rho\sigma} b_\mu \left( 3 Z_A^{1/2} V_\nu F_{\rho\sigma} +{Z_A}^{3/2} A_\nu F_{\rho\sigma}^5\right),
\end{equation}
which implies a ratio between the transversal conductivities:
\begin{equation}
\frac{\sigma_A}{\sigma_V}= \frac{1}{3}Z_A= \frac{1}{3}\left( \frac{b_{\rm IR}}{b}\right)^2.
\end{equation}
In holography we have shown that the infrared effective coupling is reproduced by the horizon value of the background axial field, so that $\frac{b_{\rm IR}}{b}=\frac{A_z(0)}{b}$. We then expect:
\begin{equation}
\frac{\sigma_A}{\sigma_V}= \frac{1}{3} \left( \frac{A_z(r_H)}{b}\right)^2, \label{prediction}
\end{equation}
while we can recover the right $1/3$ coefficient if we express the quantity as a function of the ``screened'' infrared fields $A^{\rm IR}_{M}=\sqrt{Z_A} A_{M}$.

We can now compare our prediction with the numerical results in figure \ref{cond}. We notice the remarkable fit of the prediction to the data, which still holds in the finite temperature regime, with the transition smoothed out to a crossover.

Through this interpretation we can then recover holographically both the $1/3$ coefficient, coming from the UV anomaly structure of the theory, as well as the right dependence on the infrared renormalized couplings. Thus, our study displays both the high and low energy character of this phenomenon, as should be expected from its relation to the anomalies of the theory.

\section{Quantum phase transitions from superstrings}\label{sec:topdown}

In this section we point out that the quantum phase transition also occurs in a top down model based on a consistent truncation of IIB-supergravity \cite{Liu:2010sa,Cassani:2010uw}. The interest of such a study is twofold. On one hand, it is important to justify the consistency of the top-down holographic model (\ref{action}) from a rigorous perspective, for which string theory is the perfect scenario, as many dualities have been precisely linked to the low energy limit of brane systems. On the other hand, the string theoretical realization gives a potent geometrical insight on the topological nature of the transition, as the relevant field configurations can be recovered in terms of the eleven dimensional string geometry. As a side note, the presence of this new class of RG flows at zero temperature may allow for better understanding of the dual gauge theories.

\subsection{String model}
The truncations we are going to examine are realized, in the context of type IIB string theory, by placing a stack on $N$ D3 branes at the tip of a six dimensional Calabi-Yau cone $Y_6$. In the low energy limit the spacetime geometry in the presence of the stack of branes factorizes into the product $AdS_5 \times X_5$, where $X_5$ is a Sasaki-Einstein manifold \cite{Klebanov:1998hh}:
\begin{equation}\label{stringmetric}
ds^2= ds^2_{AdS} + L^2 ds^2_{X_5}.
\end{equation}
The gauge duals of such particular configurations have been extensively analyzed in the literature and they describe $\mathcal{N}=1$ $SU(N)$ (quiver) gauge theories with a superpotential $W$. Their precise spectrum and their internal symmetries depend on the choice of the Sasaki-Einstein manifold $X_5$.
The simplest example of such construction is the case in which $X_5=T^{1,1}\approx \frac{SU(2)\times SU(2)}{U(1)}$, whose gauge dual is the $\mathcal{N}=1$, $SU(N)\times SU(N)$ Klebanov-Witten theory.
It is useful to describe the Sasaki-Einstein structure as a $U(1)$ fibration over a four dimensional Kh\"{a}ler-Einstein manifold by a globally defined vector field $\eta$ \cite{Liu:2010sa}, so that its metric takes the simple form:
\begin{equation}
ds^2_{X_5}=ds^2_{KE} + \eta \otimes \eta,
\end{equation}
and the vector field $\eta$ corresponds in this notation to the $U(1)$ R-symmetry of the dual gauge theory.

Modifications of these background geometries were studied for example in \cite{Liu:2010sa,Cassani:2010uw}. These allow the introduction of dual operator insertions that lead to interesting RG flows in the infrared. Some of these flows were already studied in \cite{Gubser:2009qm,Arean:2010wu} in the context of holographic superconductors.  
Here we focus on a different branch of solutions of the same truncation, for which a nonzero scalar source is turned on at the conformal boundary. This differs from the superconductor case, in which the source term is developed in the infrared as a consequence of the spontaneous breaking of the $U(1)$ symmetry and the gauge field represents superfluid flow.

The ten dimensional metric (\ref{stringmetric}) is in this case modified by a dynamical scalar field $\phi$ and a one form $A$ into:
\begin{equation}\label{stringmetric2}
ds^2= \cosh\left(\frac{\phi}{2}\right) ds^2_5 +\frac{L^2}{\cosh\left(\frac{\phi}{2}\right)}\left(ds^2_{KE} +\cosh^2\left(\frac{\phi}{2}\right)(\eta +\frac{2}{3}A)^2 \right)
,\end{equation}
and one can flow back to the initial Sasaki-Einstein structure if $\phi=0$. 
After dimensional reduction of the full IIB-supergravity equations (see, for example, \cite{Cassani:2010uw}), the relevant dynamics can be described through an effective five dimensional holographic action:
\begin{align}
S_g &= \frac{1}{2\kappa_5^2} \int d^5 x \sqrt{-g}\left(R  -\frac{1}{3}F^2 +\frac{2}{27}L^3 \epsilon^{MNOPQ}A_MF_{NO}F_{PQ} -\frac{1}{2}\Sigma^2 -V\right) , \label{stringaction}\\
\Sigma_{MN} &= \partial_M \phi \partial_N \phi + \sinh^2(\phi)(\partial_M \theta -2 A_M)(\partial_N \theta -2 A_N) , \\
V &= -\frac{3}{L^2}\cosh^2(\phi/2)(5-\cosh(\phi)).
\end{align}
Let us briefly comment on its field content. First, the action contains a bulk gauge field $A_M$, whose field strength we denote by $F = dA$. This is dual to the R-current $\mathcal{J}_R^\mu$ of the supersymmetric theory, given that it appears as a natural ``gauging'' of the $U(1)$ Killing field $\eta$ which describes the R-symmetry in the gravitational geometry. This symmetry is explicitly broken in the background by $\phi\neq0$ and $\theta=0$. It is also anomalous, as the gauge field appears with a Chern-Simons term:
\begin{equation}
 S_{CS}=\frac{1}{2\kappa_5^2}\frac{2}{27}L^3 \int d^5 x \sqrt{-g}\epsilon^{MNOPQ}A_MF_{NO}F_{PQ}
 ,\end{equation}
which implements a $U(1)^3$ anomaly.

Useful information can be extracted from a series expansion of the scalar potential for small $\phi$: $V(\phi)= -\frac{12}{L^2} -\frac{3}{2 L^2}\phi^2 + \frac{1}{4 L^2}\phi^4 + \mathcal{O}(\phi^6)$. The first term permits to have AdS asymptotics through the effective cosmological constant $V(0)$. The mass term tells us that $\Phi$ is dual to an operator $\mathcal{O}_\Phi$ of scaling dimension $\Delta=3$. The quartic term, furthermore, falls into the range of stability of section 2 after taking care of the difference in the normalization between (\ref{action}) and (\ref{stringaction}). In fact, we can see the potential in our previous model as a quartic truncation of the string potential, as we see in figure \ref{potential}.  Finally, the gauge coupling of $\theta$ tells us that $\mathcal{O}_\Phi$ has R-charge 2.
We thus see that the quantum numbers of $\Phi$ correspond to those of a chiral primary operator in the QFT (as $\Delta= \frac{3}{2}q_R$) and, in particular, these quantum numbers coincide with those of the superpotential $W$, which has led to conjecture their duality \cite{Gubser:2009qm}. 

The variation of the five dimensional action (\ref{stringaction}) gives the equations of motion governing the holographic dynamics:
\begin{align}
G_{MN} -\frac{2}{3}L^2\left(F_{MP}{F_N}^P+\frac{g_{MN}}{4}F^2\right) -\frac{1}{2}\Sigma_{MN} +\frac{g_{MN}}{4}\Sigma + V/2&=0, \\
\Box \phi - \frac{1}{2} \sinh(2\phi)(\partial \theta -2 A)^2 +\frac{3}{2 L^2}\left(\sinh(\phi)(5-\cosh(\phi))- 2\cosh^2(\phi/2)\sinh(\phi)\right)&=0, \\
\nabla_M F^{MN} +\frac{3}{2 L^2}\sinh^2(\phi)(\partial^N \theta- 2 A^N) +\frac{L}{6}\epsilon^{MNOPQ}F_{NO}F_{PQ}&=0,
\end{align}
which can be used to analyze the RG dynamics of the theory through holography.

\subsection{Phase transition}

We can study the zero temperature anisotropic solutions of the form with the same ansatz as before: 
\begin{equation}
ds^2= u(r)\left(-dt^2 +dx^2 +dy^2\right) + \frac{dr^2}{u(r)} + h(r) dz^2
,\end{equation}
\begin{equation}
A_M dx^M=A_z(r) dz, \ \ \phi=\phi(r), \ \ \partial_r \theta = 2 A_r
,\end{equation}
together with the boundary Anti de Sitter asymptotics and 
\begin{equation}
\lim_{r \to \infty} r \phi(r)=M, \hspace{0.5cm}  \lim_{r \to \infty}A_z=b . 
\end{equation}
In this case the introduction of a nonzero source for the R-charged scalar breaks the R-symmetry explicitly in the UV theory. As in the previous case, the system can be seen to present two different infrared phases, which are signaled by two distinct vacuum configurations that will correspond to the two infrared vacua of the theory.

\begin{figure}\resizebox{0.6\textwidth}{!}{\input{potential1.tex}} \ \ \raisebox{3cm}{\begin{tabular}{cc}
$\bar{\phi}$ & $m^2 L^2$ \\
\hline
\vspace{1mm} 
$0$ & $-3$ \\ 
 $\log(2+\sqrt{3})$ & $9$ \\ 
\hline  
\end{tabular}}
 \caption{Vacua of the string model. The blue line represents the fourth order truncation of the potential we are considering in the first sections.}
 \label{potential}
\end{figure}

From the metric (\ref{stringmetric2}) we see that only in the $\phi=0$ vacuum supersymmetry persists in the infrared, while in the other case the metric has no Killing spinor to support the $\mathcal{N}=1$ supersymmetry. Following the near horizon analysis of section \ref{sec:qpt}, we can find two distinct infrared asymptotics, which will correspond to the two phases:

\begin{description}
\item [Phase I:] $\displaystyle\lim_{r \to 0} A_z= A_z(0) \neq 0,\ \displaystyle\lim_{r \to 0} \phi=0$.\\
In this phase, the system interpolates between two supersymmetric vacua with different strengths of the coupling dual to $A_z$. The explicit breaking of the R-symmetry reduces to an anomalous breaking due to the infrared coupling $A_z(0)$ which enters the Chern-Simons term.  The presence of such a coupling triggers an anomalous transverse conductivity which survives at zero temperature, along the lines of what we have seen in section \ref{sec:sigmaaxial}. This is a signature of the  topological nature of the phase.
\item [Phase II:] $\displaystyle\lim_{r \to 0} A_z= 0,\ \displaystyle\lim_{r \to 0} \phi= \log(2+\sqrt{3})$.\\
In this case, the system breaks the supersymmetric configuration along the RG flow, of course the R symmetry is still explicitly broken and this effect leads the transverse conductivity to vanish.
\end{description}

\begin{figure}
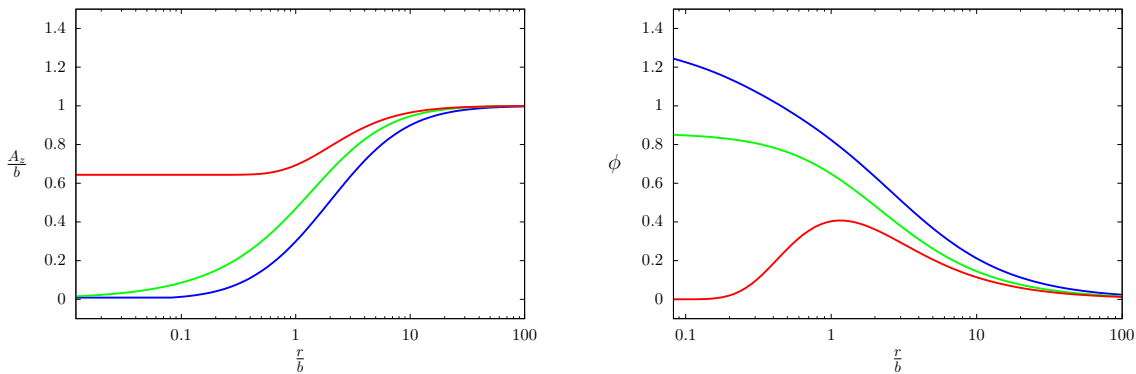

\centering
\resizebox{0.48\textwidth}{!}{\input{fieldAz1.tex}} \hfill \resizebox{0.48\textwidth}{!}{\input{fieldPhi1.tex}}
\caption{Running of the gauge field $A_z$ and the scalar field $\phi$ with the bulk coordinate $r/b$ in the three phases (topological (red), critical (green) and trivial (blue)). One can clearly see the flowing to the three different infrared asymptotics.}
\label{fields}
\end{figure}

These two branches are separated by a quantum critical point with Lifshitz-type scaling as discussed in detail in section \ref{sec:qpt}. We can continue these infrared phases to their AdS asymptotics so that each solution of the family above will be labeled by the dimensionless ratio $M/b$. The full solutions for the three cases are reported in figure \ref{fields}. Among their numerical parameters of particular importance is the anomalous scaling exponent of the critical solution $\beta$. This controls the scaling of the QFT quantities around the critical point. For it we find a numerical value of $\beta=0.821$. 

Furthermore a computation along the lines of section \ref{sec:sigmaaxial} permits to find the anomalous transversal conductivity from the Kubo formula (\ref{kubo}). In this case, the perturbative equations of motion read:
\begin{equation}
\begin{aligned}
 a_x'' + \left( \frac{h'}{2 h} + \frac{u'}{u} \right) a_x' + \frac{\omega^2}{u^2} a_x + \frac{4 i \omega A_z'}{3 u \sqrt{h}} a_y - \frac{3 \sinh^2 \phi}{u} a_x - \frac{u A_z'}{h} {h^x}_z' &= 0 , \\
 a_y'' + \left( \frac{h'}{2 h} + \frac{u'}{u} \right) a_y' + \frac{\omega^2}{u^2} a_y - \frac{4 i \omega A_z'}{3 u \sqrt{h}} a_x - \frac{3 \sinh^2 \phi}{u} a_y - \frac{u A_z'}{h} {h^y}_z' &= 0 , \\
 {h^x}_z'' + \left( 3 \frac{u'}{u} - \frac{h'}{2 h} \right) {h^x}_z' + \frac{\omega^2}{u^2} {h^x}_z + \frac{4 A_z'}{3 u} a_x' + \frac{4 A_z \sinh^2 \phi}{u^2} a_x &= 0 , \\
 {h^y}_z'' + \left( 3 \frac{u'}{u} - \frac{h'}{2 h} \right) {h^y}_z' + \frac{\omega^2}{u^2} {h^y}_z + \frac{4 A_z'}{3 u} a_y' + \frac{4 A_z \sinh^2 \phi}{u^2} a_y &= 0 ,
\end{aligned}
\end{equation}
and the $A$ and $B$ matrices in the second-order on-shell action are:
\begin{equation}\label{eq:matricesIIB}
\begin{aligned}
 A &= \left( 
 \begin{array}{c c c c}
  -\frac{2 \sqrt{h} u}{3} & 0 & 0 & 0 \\
  0 & - \frac{u^3}{2 \sqrt{h}} & 0 & 0 \\
  0 & 0 & -\frac{2 \sqrt{h} u}{3} & 0 \\
  0 & 0 & 0 & - \frac{u^3}{2 \sqrt{h}} \\
 \end{array}
 \right),\\\\
 B &= \left( 
 \begin{array}{c c c c}
  0 & \frac{4 u^2 A_z'}{3 \sqrt{h}} & - \frac{16 i \omega \alpha A_z}{27} & 0 \\
  0 & - \frac{3 u^2 u'}{\sqrt{h}} & 0 & 0 \\
  \frac{16 i \omega \alpha A_z}{27} & 0 & 0 & \frac{4 u^2 A_z'}{3 \sqrt{h}} \\
  0 & 0 & 0 & - \frac{3 u^2 u'}{2 \sqrt{h}} \\
 \end{array}
 \right).
\end{aligned}
\end{equation}

Based on the charge screening arguments of section 3 we would expect the transversal conductivity in the R-current to be:
\begin{equation}
\sigma_R = \frac{16}{27} \frac{L^8}{2\kappa_5^2} \frac{A_z(0)^3}{b^2}= \frac{N^2}{8 \pi^2}\sqrt{Z_A^3} b , \label{sugrapred}
\end{equation}
where again we have  defined the renormalization constant charge as $\sqrt{Z_A} = \frac{A_z(0)}{b}$. 

\begin{figure}
\centering
\resizebox{0.6\textwidth}{!}{\input{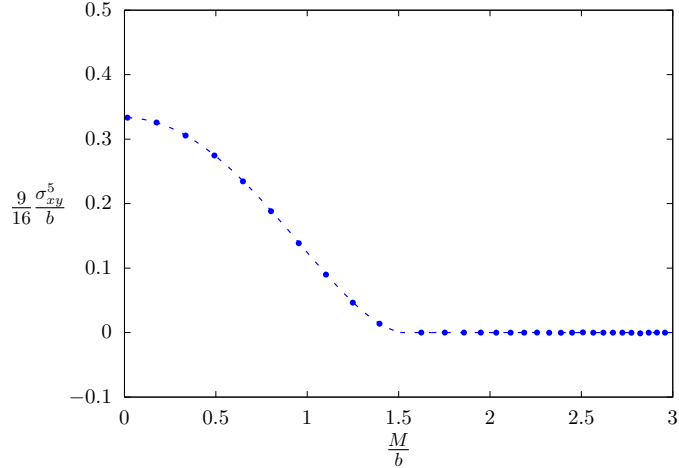}}
\caption{$U(1)_R^3$ anomalous transverse conductivity, normalized to the UV anomaly coefficient, computed from the string action (\ref{stringaction}) as a function of $M/b$. The dashed line denote the prediction (\ref{sugrapred}) while dots stand for the numerical values. Notice again the agreement with the prediction}
\label{stringcond}
\end{figure}

The results, that can be seen in figure \ref{stringcond}, confirm the consistency of our intuition and give a natural order parameter for the SUSY-breaking quantum phase transition described in this section.

\section{Conclusions}
We have studied the dependence of the quantum phase transition of the holographic Weyl-semimetal state on the parameters
of the scalar potential in a bottom-up model. We found that generically it persists but that the trivial phase becomes 
unaccessible for large quartic scalar self coupling or close to the marginality bound on the dual operator. 
We also found that the phase transition is present in a top-down string theory based model.

We have computed the axial Hall conductivity and found that after taking a non-trivial renormalization of the axial gauge
field into account it is precisely $1/3$ of the electric hall conductivity. This is indeed what can be expected from
general arguments based on the algebraic structure of the axial anomaly. While on a fundamental level axial gauge fields 
are not present in nature they can appear as effects of strain in Weyl semi-metals \cite{Cortijo:2016wnf,axialmagneticAdolfo,Pikulin:2016wfj}.

\section{Acknowledgments}
We would like to thank Y. Liu for useful discussions and help on the numerics.
This research has been supported by  FPA2015-65480-P and by the Centro de 
Excelencia Severo Ochoa Programme under grant SEV-2012-0249. The work of
J.F.-P. is supported by fellowship SEV-2012-0249-03. The work of C.C. is funded 
by Fundación La Caixa under ``La Caixa-Severo Ochoa'' international predoctoral grant.

\bibliography{AnomTrans}{}

\providecommand{\href}[2]{#2}\begingroup\raggedright\begin{thebibliography}{10}

\bibitem{WSM1}
X.~Wan, A.~M. Turner, A.~Vishwanath and S.~Y. Savrasov, \emph{Topological
  semimetal and fermiarc surface states in the electronic structure of
  pyrochlore iridates}, {\emph{Phys. Rev.} {\bf B 83} (2011) 205101},
  [\href{https://arxiv.org/abs/1007.0016}{{\tt 1007.0016}}].

\bibitem{WSMreviewTV}
A.~V. Ari M.~Turner, \emph{{Beyond Band Insulators: Topology of Semi-metals and
  Interacting Phases}},  \href{https://arxiv.org/abs/1301.0330}{{\tt
  1301.0330}}.

\bibitem{Landsteiner:2016led}
K.~Landsteiner, \emph{{Notes on Anomaly Induced Transport}},  in \emph{{56th
  Cracow School of Theoretical Physics: A Panorama of Holography Zakopane,
  Poland, May 24-June 1, 2016}}, 2016.
\newblock \href{https://arxiv.org/abs/1610.04413}{{\tt 1610.04413}}.

\bibitem{Fukushima:2008xe}
K.~Fukushima, D.~E. Kharzeev and H.~J. Warringa, \emph{{The Chiral Magnetic
  Effect}}, \href{http://dx.doi.org/10.1103/PhysRevD.78.074033}{\emph{Phys.
  Rev.} {\bf D78} (2008) 074033}, [\href{https://arxiv.org/abs/0808.3382}{{\tt
  0808.3382}}].

\bibitem{Li:2014bha}
Q.~Li, D.~E. Kharzeev, C.~Zhang, Y.~Huang, I.~Pletikosic, A.~V. Fedorov et~al.,
  \emph{{Observation of the chiral magnetic effect in ZrTe5}},
  \href{http://dx.doi.org/10.1038/nphys3648}{\emph{Nature Phys.} {\bf 12}
  (2016) 550--554}, [\href{https://arxiv.org/abs/1412.6543}{{\tt 1412.6543}}].

\bibitem{Landsteiner:2015pdh}
K.~Landsteiner, Y.~Liu and Y.-W. Sun, \emph{{Quantum phase transition between a
  topological and a trivial semimetal from holography}},
  \href{http://dx.doi.org/10.1103/PhysRevLett.116.081602}{\emph{Phys. Rev.
  Lett.} {\bf 116} (2016) 081602},
  [\href{https://arxiv.org/abs/1511.05505}{{\tt 1511.05505}}].

\bibitem{Landsteiner:2016stv}
K.~Landsteiner, Y.~Liu and Y.-W. Sun, \emph{{Odd viscosity in the quantum
  critical region of a holographic Weyl semimetal}},
  \href{http://dx.doi.org/10.1103/PhysRevLett.117.081604}{\emph{Phys. Rev.
  Lett.} {\bf 117} (2016) 081604},
  [\href{https://arxiv.org/abs/1604.01346}{{\tt 1604.01346}}].

\bibitem{Grignani:2016wyz}
G.~Grignani, A.~Marini, F.~Pena-Benitez and S.~Speziali, \emph{{AC conductivity
  for a holographic Weyl Semimetal}},
  \href{https://arxiv.org/abs/1612.00486}{{\tt 1612.00486}}.

\bibitem{Ammon:2016mwa}
M.~Ammon, M.~Heinrich, A.~Jiménez-Alba and S.~Moeckel, \emph{{Surface States
  in Holographic Weyl Semimetals}},
  \href{https://arxiv.org/abs/1612.00836}{{\tt 1612.00836}}.

\bibitem{Jackiw:1999qq}
R.~Jackiw, \emph{{When radiative corrections are finite but undetermined}},
  \href{http://dx.doi.org/10.1142/S021797920000114X}{\emph{Int. J. Mod. Phys.}
  {\bf B14} (2000) 2011--2022},
  [\href{https://arxiv.org/abs/hep-th/9903044}{{\tt hep-th/9903044}}].

\bibitem{Landsteiner:2015lsa}
K.~Landsteiner and Y.~Liu, \emph{{The holographic Weyl semi-metal}},
  \href{http://dx.doi.org/10.1016/j.physletb.2015.12.052}{\emph{Phys. Lett.}
  {\bf B753} (2016) 453--457}, [\href{https://arxiv.org/abs/1505.04772}{{\tt
  1505.04772}}].

\bibitem{Gursoy:2012ie}
U.~Gursoy, V.~Jacobs, E.~Plauschinn, H.~Stoof and S.~Vandoren,
  \emph{{Holographic models for undoped Weyl semimetals}},
  \href{http://dx.doi.org/10.1007/JHEP04(2013)127}{\emph{JHEP} {\bf 04} (2013)
  127}, [\href{https://arxiv.org/abs/1209.2593}{{\tt 1209.2593}}].

\bibitem{Jacobs:2015fiv}
V.~P.~J. Jacobs, P.~Betzios, U.~Gursoy and H.~T.~C. Stoof,
  \emph{{Electromagnetic response of interacting Weyl semimetals}},
  \href{http://dx.doi.org/10.1103/PhysRevB.93.195104}{\emph{Phys. Rev.} {\bf
  B93} (2016) 195104}, [\href{https://arxiv.org/abs/1512.04883}{{\tt
  1512.04883}}].

\bibitem{Hashimoto:2016ize}
K.~Hashimoto, S.~Kinoshita, K.~Murata and T.~Oka, \emph{{Holographic Floquet
  states: (I) A strongly coupled Weyl semimetal}},
  \href{https://arxiv.org/abs/1611.03702}{{\tt 1611.03702}}.

\bibitem{Cortijo:2016yph}
A.~Cortijo, Y.~Ferreiros, K.~Landsteiner and M.~A.~H. Vozmediano,
  \emph{{Elastic Gauge Fields in Weyl Semimetals}},
  \href{http://dx.doi.org/10.1103/PhysRevLett.115.177202}{\emph{Phys. Rev.
  Lett.} {\bf 115} (2015) 177202},
  [\href{https://arxiv.org/abs/1603.02674}{{\tt 1603.02674}}].

\bibitem{Arean:2010wu}
D.~Arean, M.~Bertolini, C.~Krishnan and T.~Prochazka, \emph{{Type IIB
  Holographic Superfluid Flows}},
  \href{http://dx.doi.org/10.1007/JHEP03(2011)008}{\emph{JHEP} {\bf 03} (2011)
  008}, [\href{https://arxiv.org/abs/1010.5777}{{\tt 1010.5777}}].

\bibitem{Girardello:1998pd}
L.~Girardello, M.~Petrini, M.~Porrati and A.~Zaffaroni, \emph{{Novel local CFT
  and exact results on perturbations of N=4 superYang Mills from AdS
  dynamics}},
  \href{http://dx.doi.org/10.1088/1126-6708/1998/12/022}{\emph{JHEP} {\bf 12}
  (1998) 022}, [\href{https://arxiv.org/abs/hep-th/9810126}{{\tt
  hep-th/9810126}}].

\bibitem{Freedman:1999gp}
D.~Z. Freedman, S.~S. Gubser, K.~Pilch and N.~P. Warner, \emph{{Renormalization
  group flows from holography supersymmetry and a c theorem}}, {\emph{Adv.
  Theor. Math. Phys.} {\bf 3} (1999) 363--417},
  [\href{https://arxiv.org/abs/hep-th/9904017}{{\tt hep-th/9904017}}].

\bibitem{Klebanov:2002gr}
I.~R. Klebanov, P.~Ouyang and E.~Witten, \emph{{A Gravity dual of the chiral
  anomaly}}, \href{http://dx.doi.org/10.1103/PhysRevD.65.105007}{\emph{Phys.
  Rev.} {\bf D65} (2002) 105007},
  [\href{https://arxiv.org/abs/hep-th/0202056}{{\tt hep-th/0202056}}].

\bibitem{Casero:2007ae}
R.~Casero, E.~Kiritsis and A.~Paredes, \emph{{Chiral symmetry breaking as open
  string tachyon condensation}},
  \href{http://dx.doi.org/10.1016/j.nuclphysb.2007.07.009}{\emph{Nucl. Phys.}
  {\bf B787} (2007) 98--134}, [\href{https://arxiv.org/abs/hep-th/0702155}{{\tt
  hep-th/0702155}}].

\bibitem{Gursoy:2014ela}
U.~Gürsoy and A.~Jansen, \emph{{(Non)renormalization of Anomalous
  Conductivities and Holography}},
  \href{http://dx.doi.org/10.1007/JHEP10(2014)092}{\emph{JHEP} {\bf 10} (2014)
  092}, [\href{https://arxiv.org/abs/1407.3282}{{\tt 1407.3282}}].

\bibitem{Jimenez-Alba:2014iia}
A.~Jimenez-Alba, K.~Landsteiner and L.~Melgar, \emph{{Anomalous magnetoresponse
  and the Stückelberg axion in holography}},
  \href{http://dx.doi.org/10.1103/PhysRevD.90.126004}{\emph{Phys. Rev.} {\bf
  D90} (2014) 126004}, [\href{https://arxiv.org/abs/1407.8162}{{\tt
  1407.8162}}].

\bibitem{Grushin:2012mt}
A.~G. Grushin, \emph{{Consequences of a condensed matter realization of Lorentz
  violating QED in Weyl semi-metals}},
  \href{http://dx.doi.org/10.1103/PhysRevD.86.045001}{\emph{Phys. Rev.} {\bf
  D86} (2012) 045001}, [\href{https://arxiv.org/abs/1205.3722}{{\tt
  1205.3722}}].

\bibitem{Iqbal:2008by}
N.~Iqbal and H.~Liu, \emph{{Universality of the hydrodynamic limit in AdS/CFT
  and the membrane paradigm}},
  \href{http://dx.doi.org/10.1103/PhysRevD.79.025023}{\emph{Phys. Rev.} {\bf
  D79} (2009) 025023}, [\href{https://arxiv.org/abs/0809.3808}{{\tt
  0809.3808}}].

\bibitem{Cortijo:2016wnf}
A.~Cortijo, D.~Kharzeev, K.~Landsteiner and M.~A.~H. Vozmediano, \emph{{Strain
  induced Chiral Magnetic Effect in Weyl semimetals}},
  \href{https://arxiv.org/abs/1607.03491}{{\tt 1607.03491}}.

\bibitem{Pikulin:2016wfj}
D.~I. Pikulin, A.~Chen and M.~Franz, \emph{{Chiral anomaly from strain-induced
  gauge fields in Dirac and Weyl semimetals}},
  \href{https://arxiv.org/abs/1607.01810}{{\tt 1607.01810}}.

\bibitem{axialmagneticAdolfo}
A.~V. Adolfo G.~Grushin, Jorn W. F.~Venderbos and R.~Ilan, \emph{{Inhomogeneous
  Weyl and Dirac semimetals: Transport in axial magnetic fields and Fermi arc
  surface states from pseudo Landau levels}}, {\emph{.} (2016) },
  [\href{https://arxiv.org/abs/1607.04268}{{\tt 1607.04268}}].

\bibitem{Kaminski:2009dh}
M.~Kaminski, K.~Landsteiner, J.~Mas, J.~P. Shock and J.~Tarrio,
  \emph{{Holographic Operator Mixing and Quasinormal Modes on the Brane}},
  \href{http://dx.doi.org/10.1007/JHEP02(2010)021}{\emph{JHEP} {\bf 02} (2010)
  021}, [\href{https://arxiv.org/abs/0911.3610}{{\tt 0911.3610}}].

\bibitem{Goswami:2012db}
P.~Goswami and S.~Tewari, \emph{{Axionic field theory of (3+1)-dimensional Weyl
  semimetals}}, \href{http://dx.doi.org/10.1103/PhysRevB.88.245107}{\emph{Phys.
  Rev.} {\bf B88} (2013) 245107}, [\href{https://arxiv.org/abs/1210.6352}{{\tt
  1210.6352}}].

\bibitem{Zyuzin:2012tv}
A.~A. Zyuzin and A.~A. Burkov, \emph{{Topological response in Weyl semimetals
  and the chiral anomaly}},
  \href{http://dx.doi.org/10.1103/PhysRevB.86.115133}{\emph{Phys. Rev.} {\bf
  B86} (2012) 115133}, [\href{https://arxiv.org/abs/1206.1868}{{\tt
  1206.1868}}].

\bibitem{Liu:2010sa}
J.~T. Liu, P.~Szepietowski and Z.~Zhao, \emph{{Consistent massive truncations
  of IIB supergravity on Sasaki-Einstein manifolds}},
  \href{http://dx.doi.org/10.1103/PhysRevD.81.124028}{\emph{Phys. Rev.} {\bf
  D81} (2010) 124028}, [\href{https://arxiv.org/abs/1003.5374}{{\tt
  1003.5374}}].

\bibitem{Cassani:2010uw}
D.~Cassani, G.~Dall'Agata and A.~F. Faedo, \emph{{Type IIB supergravity on
  squashed Sasaki-Einstein manifolds}},
  \href{http://dx.doi.org/10.1007/JHEP05(2010)094}{\emph{JHEP} {\bf 05} (2010)
  094}, [\href{https://arxiv.org/abs/1003.4283}{{\tt 1003.4283}}].

\bibitem{Klebanov:1998hh}
I.~R. Klebanov and E.~Witten, \emph{{Superconformal field theory on
  three-branes at a Calabi-Yau singularity}},
  \href{http://dx.doi.org/10.1016/S0550-3213(98)00654-3}{\emph{Nucl. Phys.}
  {\bf B536} (1998) 199--218},
  [\href{https://arxiv.org/abs/hep-th/9807080}{{\tt hep-th/9807080}}].

\bibitem{Gubser:2009qm}
S.~S. Gubser, C.~P. Herzog, S.~S. Pufu and T.~Tesileanu, \emph{{Superconductors
  from Superstrings}},
  \href{http://dx.doi.org/10.1103/PhysRevLett.103.141601}{\emph{Phys. Rev.
  Lett.} {\bf 103} (2009) 141601}, [\href{https://arxiv.org/abs/0907.3510}{{\tt
  0907.3510}}].

\end{thebibliography}\endgroup
\bibliographystyle{JHEP}

\end{document}